\documentclass[review]{elsarticle}
\makeatletter
\def\ps@pprintTitle{%
 \let\@oddhead\@empty
 \let\@evenhead\@empty
 \def\@oddfoot{}%
 \let\@evenfoot\@oddfoot}
\makeatother

\usepackage{lineno}
\modulolinenumbers[5]
\usepackage{mathtools}
\usepackage{bm}% bold math
\usepackage{mathptmx}
\usepackage{amssymb}

\usepackage{newtxtext,newtxmath}
\usepackage{hyperref}
\usepackage{textcomp}
\usepackage{subfigure}
\usepackage[ruled,vlined]{algorithm2e}
\usepackage{dcolumn}% Align table columns on decimal point
\usepackage{amsmath}
\usepackage{xcolor}
\usepackage{graphicx}

\let\linenumbers\nolinenumbers\nolinenumbers
\begin{document}

\begin{frontmatter}

\title{Learning black- and gray-box chemotactic PDEs/closures\\
from agent based Monte Carlo simulation data}

% \tnotetext[mytitlenote]{Fully documented templates are available in the elsarticle package on \href{http://www.ctan.org/tex-archive/macros/latex/contrib/elsarticle}{CTAN}.}

% \address{Radarweg 29, Amsterdam}
% \fntext[myfootnote]{Since 1880.}

% \author[mymainaddress,mysecondaryaddress]{Elsevier Inc}

\author[sjsu]{Seungjoon Lee \textsuperscript{\textdagger}\footnote{ \textsuperscript{\textdagger}  The authors contribute equally to this paper.}}
\author[jhu]{Yorgos M. Psarellis \textsuperscript{\textdagger} }
\author[napoly]{Constantinos I. Siettos} 
\author[jhu,jhu1,jhu2]{Ioannis G. Kevrekidis\corref{mycorrespondingauthor}}
\cortext[mycorrespondingauthor]{Corresponding author}
\ead{yannisk@jhu.edu}
\date{}

\address[sjsu]{Department of Applied Data Science, San Jos{\'e} State University}
\address[jhu]{Department of Chemical and Biomolecular Engineering, Johns Hopkins University}
\address[napoly]{Dipartimento di Matematica e Applicazioni ``Renato Caccioppoli" \& Scuola Superiore Meridionale, Universit\'a degli Studi di Napoli Federico II}
\address[jhu1]{Department of Applied Mathematics and Statistics, Johns Hopkins University}
\address[jhu2]{Department of Medicine, Johns Hopkins University}

\begin{abstract}
We propose a machine learning framework for the data-driven discovery
of macroscopic chemotactic Partial Differential Equations (PDEs) 
--and the closures that lead to them-  
from high-fidelity, individual-based stochastic simulations of {\em E.coli} bacterial motility.
The fine scale, detailed, hybrid (continuum - Monte Carlo) simulation model embodies the underlying biophysics, and its parameters are informed from experimental observations of individual cells.
We exploit Automatic Relevance Determination (ARD) within a Gaussian Process framework for the identification of a parsimonious set of collective observables that parametrize {\em the law} of the effective PDEs.
Using these observables, in a second step we learn effective, coarse-grained ``Keller-Segel class" chemotactic PDEs using machine learning regressors: (a) (shallow) feedforward neural networks and (b) Gaussian Processes. 
The  learned laws can be {\em black-box} (when no prior knowledge about the PDE law structure is assumed) or {\em gray-box} when parts of the equation (e.g. the pure diffusion part) is known and ``hardwired" in the regression process.
We also discuss data-driven corrections (both additive and functional) of analytically known, approximate closures. 

\end{abstract}

%======================================================================
%Subjects for ARXIV: Artificial Intelligence (cs.AI); Dynamical Systems (math.DS); Numerical Analysis (math.NA); Machine Learning (stat.ML); 
%======================================================================

\end{frontmatter}

\linenumbers

\section{Introduction}

Since the pioneering work of Adler \cite{adler1969chemoreceptors}, the chemotactic motility of bacteria, i.e., their movement in response to changes in the surrounding environment has been extensively studied, thus decoding complex mechanisms ranging from the biochemistry and molecular genetics \cite{parkinson1976chea,parkinson1980novel} to the inter \cite{boyd1981sensory} and intracellular signaling \cite{liu1989role,heit2002intracellular} and from the sensory adaptation in response to external stimuli \cite{segel1986mechanism,othmer1998oscillatory} to the motor structure and the flagellum-related motility \cite{cluzel2000ultrasensitive} and scaling up to the emergent collective behaviour \cite{wu2006collective}.
Depending on the level of information and spatio-temporal scale of analysis, a vast number of mathematical models have been proposed ranging from the molecular/individual \cite{emonet2005agentcell,coburn2013tactile,othmer2013excitation,rousset2013simulating,yasuda2017monte} to the continuum/macroscopic scale \cite{patlak1953mathematical,keller1971,erban2004,bellomo2010multiscale,othmer2013excitation,franz2013hybrid,bellomo2022chemotaxis} (for an extensive review of both modelling approaches see  \cite{tindall2008overview1,tindall2008overview2,othmer2013excitation,bellomo2022chemotaxis}.
The celebrated Keller-Segel~\cite{keller1971} PDE derived for the macroscopic description of the population density evolution, coupled with the concommitant chemoattractant field, constitutes the cornerstone in the field. In its simplest form, the dependence of the cell density $b(x,t)$ in space and time evolves according to
\begin{equation}
    \frac{\partial b}{\partial t} = \nabla \cdot \left( D\nabla b - \chi(s) b \nabla s \right),
    \label{eqn:KS}
\end{equation}
coupled with appropriate boundary conditions; the diffusion term in general depends on the non-chemotactic random motility of the bacteria; here, $D$ is the diffusion coefficient and $\chi$ is the {\em chemotactic coefficient}. 
%The Keller-Segel PDE is derived from kinetic theory based on an approximation used in describing collective dynamics of Brownian motion. 
In general, the effect of the substrate's distribution and the population density on both $D$ and $\chi$ are not known. In order to obtain an expression in closed-form and then attempt to fuse experimental observations and bio-physical insight, several assumptions are made, resulting to different closed-form approximations. 
For example, in the original paper of Keller and Segel, \cite{keller1971} it is assumed that $D=D(s)$ and $\chi=\chi(s)$, i.e., that both depend on the distribution of the substrate $s$ (in general $s=s(x,t)$).
In their more general forms, the diffusion and chemotactic coefficients depend on both the density $b$ and the substrate profile $s$, i.e. $D=D(b,s)$, $\chi=\chi(b,s)$. Assuming $D=D(b)$, $\chi=\chi(b)$, the Keller-Segel PDE reduces to a Fokker-Planck equation, while for constant diffusion $D$ and constant chemotactic $\chi$ coefficient, we obtain the Smoluchowski equation (for a review of different closures and models refer to \cite{othmer1998oscillatory,chavanis2008nonlinear,erban2004, othmer2013excitation,painter2019mathematical}).
Obtaining an accurate constitutive relation (a closure) for the diffusivity $D$ and the chemotactic sensitivity $\chi$, one that matches a specific experimental setup and/or using statistical mechanics, starting from cell-based models, remains both a challenging and open-ended research task.
The goal is to obtain closed-form PDEs, such as those based on the Patlak-Keller-Segel (PKS) model \cite{patlak1953mathematical,erban2004,kim2012patlak,othmer2013excitation} that can efficiently describe the observed collective dynamics.\par
\begin{figure}[h]
  \makebox[\textwidth][c]{\includegraphics[width=1.2\textwidth]{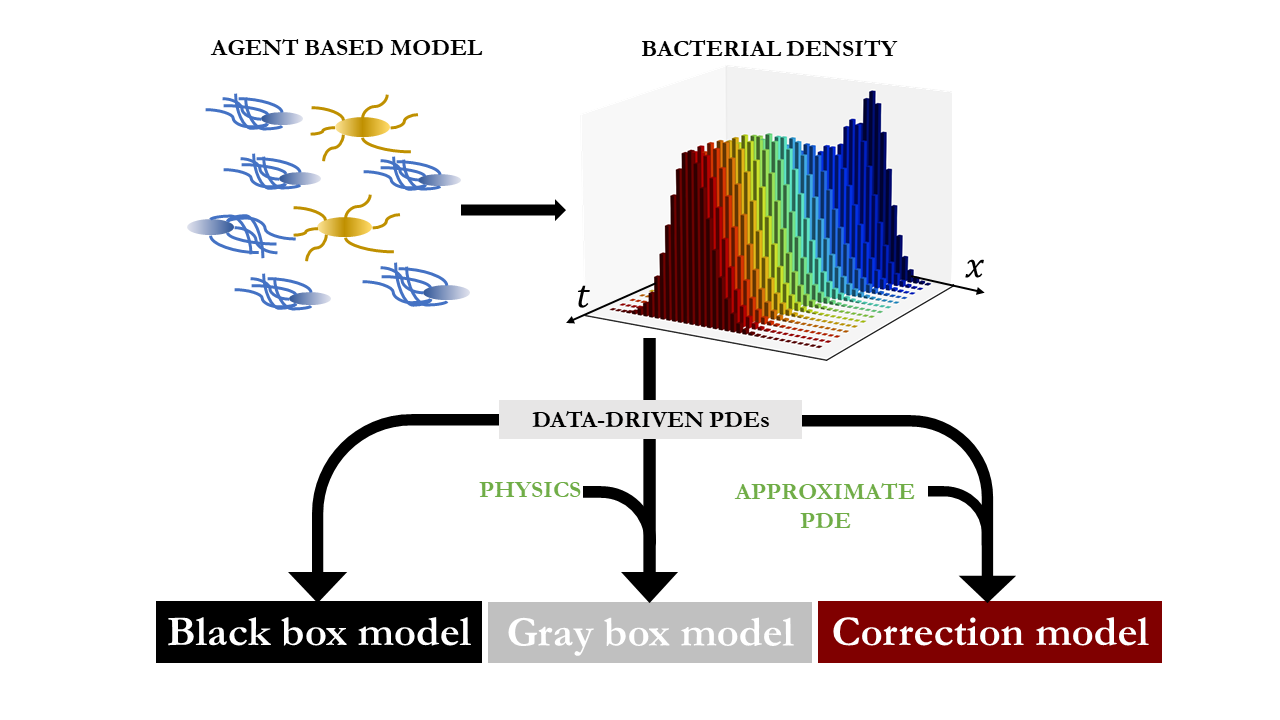}}%
  \caption{Overview of proposed algorithmic pipeline: from Agent Based Modeling (Monte Carlo simulations) to Data-driven, possibly partially physics-informed, surrogate models for Chemotactic Partial Differential Equations (PDEs).}
  \label{fig:overview}
\end{figure}

Numerical ``closure on demand" approaches, based on the Equation-free multiscale framework, that bypass the need to extract explicit macrosocpic PDEs in a closed form have been also proposed for the scientific computation of the collective motility dynamics \cite{setayeshgar2005application,erban2006equation,siettos2014}.
Here, starting from high-fidelity data, produced by a detailed realistic biophysical Monte-Carlo model for the motility of {\em E. coli} bacteria, whose parameters are calibrated from experimental data \cite{berg1972chemotaxis,larsen1974change,maeda1976effect,block1983adaptation,BLOCK1982215,ishihara1983coordination,spiro1997,othmer1998oscillatory,cluzel2000ultrasensitive,othmer2013excitation}, and building on previous efforts \cite{Rico94,Krischer93,Gonzalez98,bertalan2019learning,lee2020,kemeth2020learning}, we propose a data-based, machine-learning assisted framework to learn the law of the underlying macrosopic PDE. In particular, based on automatic relevance determination (ARD) \cite{mackay1992bayesian,sandhu2017bayesian} within the Gaussian Process Regression framework \cite{Rasmussen2005,lee2020} for feature extraction, and on Gaussian Processes and feedforward neural networks to learn the collective dynamics, we (see Fig.\ref{fig:overview}) : (a) identify the right-hand-side (RHS) of an effective Keller-Segel-class PDE, thus obtaining a {\em black-box} PDE model; (b) reconstruct the chemotactic term only, assuming that the diffusion term can be estimated by the high-fidelity simulations and/or knowledge of the physics, thus constructing a {\em gray box} Keller-Segel-class PDE model; and importantly, (c) discover data-driven corrections to established approximate closure approximations of the chemotactic term, which have been derived analytically from kinetic theory/statistical mechanics based on series of assumptions \cite{erban2004}.

The remainder of the paper is organized as follows: in section~\ref{sec:agent}, we briefly present the bio-mechanical based Monte-Carlo model and discuss its parametrization based on experimental studies.
In section~\ref{sec:method}, we present an overview of our proposed numerical framework and briefly review theoretical concepts of Gaussian Process Regression, ARD feature selection and Artificial Neural Networks.
In section~\ref{sec:models}, we demonstrate the effectiveness of our proposed framework by constructing different PDE models, namely (1) a black-box, (2) a gray-box, and (3) a model with a corrected closure for a Keller-Segel-class  PDE.
In section \ref{sec:conclusion}, we summarize our results and discuss open issues for further development of the data-driven discovery of the underlying macroscopic PDE from data generated by microscopic simulations and/or experiments.

\section{Chemotactic motility of E.coli: the bio-mechanical based Monte-Carlo model and the closed form Keller-Segel (Fokker-Planck) PDE}
\subsection{Monte-Carlo Model}
\label{sec:agent}
Here, we used a previously derived bio-mechanical-based Monte-Carlo dynamical model ~\cite{othmer2013excitation} to generate data for the chemotactic motility of {\em E.Coli} bacteria in response to a fixed chemoattractant substrate profile. 
The model calculates the probability of rotational directionality of each one of the six flagellae that extend from the surface of the cell based on changes of the concentration of the CheY-P protein, which binds to the protein FliM at the base of the rotor. The changes in the concentration of the CheY-P protein control the direction of the flagellar rotation ~\cite{othmer2013excitation,sarkar2010chemotaxis}, which in turn governs the motility of the cells. When the majority of the flagellar filaments rotate in the counter-clockwise (CCW) direction the cell swims; otherwise it tumbles. The change in time of the concentration of CheY-P protein, say $C(t)$ is represented by a simple algebraic relation reading (\cite{setayeshgar2005application, siettos2014}): $C(t)= \bar{C} -g u_1(t)$, where the dynamics of $u_1(t)$ are modeled by a simple two dimensional excitation/adaptation cartoon model given by:
\begin{equation}
    \frac{du_1}{dt}=\frac{f(s)-(u_1+u_2)}{\tau_e},\quad \frac{du_2}{dt}=\frac{f(s)-u_2}{\tau_a}, \quad f(s) = k\frac{s}{K_s+s}.
    \label{eqn:cartoon1}
\end{equation}
In the above model, $\bar{C}=2.95\mu M$ is the baseline concentration corresponding to the non-excited state ( \cite{cluzel2000ultrasensitive,setayeshgar2005application}), $g=5$ is the amplification response to excitation (\cite{setayeshgar2005application,othmer2013excitation}, $s$ represents the external stimulus (here, a chemoattractant substrate), $t_e=0.1$ and $t_a=20$ represent the excitation and adaptation time constants, respectively, and $f$ is the function encoding the signal transduction reflecting the fraction of receptors that are occupied \cite{othmer2013excitation,erban2004}; $k=15$ is a constant that amplifies the input signal and $K_s=1\mu$M is the dissociation constant for the enzyme-substrate complex \cite{erban2004,block1983adaptation,setayeshgar2005application}. The swimming speed ($v$) also depends on various parameters such as the bacteria strain, the substrate, temperature and density of cells and may vary from $\sim$ 10 to $\sim$55$\mu m /s$ \cite{berg1972chemotaxis,maeda1976effect}. For our simulations, we have set $v=30 \mu m/s$. Finally, we note that in the MC model, the movement of the bacteria is not affected by their density (they are ``noninteracting"). More details on the MC model and a pseudo-code for its implementation can be found in the Appendix and in \cite{setayeshgar2005application,siettos2014}.
\subsection{The closed form Keller-Segel PDE}
\label{sec:othmer}
For an analogous to the above microscopic description of the {\em E.coli} motility model, Erban and Othmer \cite{erban2004} with the aid of statistical mechanics/ kinetic theory, and assuming that the the signal $s(x)$ is a time independent scalar function, have derived the following parabolic Keller-Segel-class PDE in closed form in 1D:
\begin{equation}
    \frac{\partial b}{\partial t} = \frac{\partial}{\partial x}\left(\frac{\bar{v}^2}{2\lambda_0}\frac{\partial b}{\partial x} -\frac{df}{ds}\frac{c\bar{v}^2t_a}{\lambda_0(1+2\lambda_0t_a)(1+2\lambda_0t_e)}\frac{ds}{dx}b \right).
    \label{eqn:macroPDEOthmer}
\end{equation}
Here, $b=b(x,t)$ is the density of the population at $x$ and time $t$, $\bar{v}$ is the \textit{mean} speed of the bacterium's motion, $\lambda_0$ represents the basal turning frequency (frequency of tumbles) in the absence of excitation, and $c$ is a positive constant parameter that amplifies the excitation signal $u_1$ that governs the switching frequency in the presence of a stimulus (for a detailed description of the derivation of the above PDE and its parameters please refer to ~\cite{erban2004}). Based on the above, we define the \textit{generalized CHemotactic term} $CH_g$ as
\begin{equation}
CH_g \equiv  \frac{\partial}{\partial x}(\frac{df}{ds}\frac{c\bar{v}^2t_a}{\lambda_0(1+2\lambda_0t_a)(1+2\lambda_0t_e)}\frac{ds}{dx}b);
\end{equation}
the subscript $g$ refers to the generalized Keller-Segel PDE.
The main assumptions made for the above closed form PDE are the following: (a) all bacteria are running with a constant \textit{velocity} $\bar{v}$, without colliding, (b) the tumble phase is neglected, (c) the internal excitation-adaptation dynamics are described by the 2D cartoon model, (d) the distribution of the substrate $s(x)$ is a time-independent scalar function, (e) the turning frequency (frequency of tumbles in the presence of stimulus), say, $\lambda(t)$ is a linear function of $u_1(t)$ given by
\begin{equation}
   \lambda(t)=\lambda_0-c u_1(t),
   \label{eqn:lambda}
\end{equation}
and finally that (f) the gradient $s'(x)$ is shallow, so that  for all practical purposes, the second order moment of the microscopic flux is zero, i.e. that \cite{erban2004}:
\begin{equation}
   j_2(x,t)=\int_{\rm I\!R} \bar{v} \cdot (p_{+} (x,z_2,t)-p_{-} (x,z_2,t)) dz_{2} =0.
\end{equation}
$p_{\pm} (x,z_2,t)$ is the density function of the bacteria at $(x,t)$ with the internal state $z_2(x,t)=u_2(t)-s(x)$ that run right ($+$) or left ($-$). The above assumptions result to the following closed form for the chemotactic coefficient $\chi$ \cite{erban2004}:
\begin{equation}
    \chi=\frac{df}{ds}\frac{c\bar{v}^2t_a}{\lambda_0(1+2\lambda_0t_a)(1+2\lambda_0t_e) .}
\end{equation}
\section{Machine-learning and the discovery of chemotactic laws: Data-driven identification of coarse-scale PDEs}
\label{sec:method}
\subsection{Overview}
% (see Fig.\ref{fig:overview})
Before discovering effective PDE laws from microscopic simulations, there is a crucial prerequisite: what are the macroscopic observables whose field evolution laws we want to discover? There are cases for which this knowledge is given: in our case we know we want to derive parabolic evolution laws for the bacterial density field. 
Yet this knowledge is not always {\em a priori} given, based on domain knowledge: for the chemotaxis problem itself, we know that in different parameter regimes one needs a hyperbolic (higher order) equation for the density field \cite{erban2004}. 
Discovering sets of macroscopic observables in terms of which an evolutionary PDE can be closed is a nontrivial task; often this task can be performed using data mining/manifold learning techniques, and has been named ``variable-free computation" (in the sense that 
the relevant variables are identified through, say, PCA or Diffusion Map processing of the fine scale simulations (\cite{erban2007variable, arbabi2021particles}).

In our case (with the chemoattractant field $s(x)$ fixed, and with Keller-Segel-class equations in mind) we know that we want to identify a parabolic evolutionary PDE in terms of the 
evolution of a normalized bacterial density $b(x,t)$ in one spatial dimension. 
That is, we expect $\frac{\partial b}{\partial t} = F\left(b,\frac{\partial b}{\partial x}, \frac{\partial^2 b}{\partial x^2}, \dots, s, \frac{d s}{d x}, \frac{d^2 s}{dx^2}, \dots \right)$.
The existence of such a relation between the local bacterial density time derivative and the local
values and spatial derivatives of this field, as well as of the chemoattractant field, is our working hypothesis.
The question then becomes: how many spatial derivatives of the $b(x,t), s(x,t)$ fields are required in order to infer a useful data-driven closure ? 
Starting with an assumed highest order of possibly influential spatial derivatives, we resolve this here using a feature selection method. 
More specifically, we use the automatic relevance determination (ARD) in the Gaussian framework~\cite{Rasmussen2005}, which has been widely used to identify dominant input features for a certain target output using sensitivity analysis~\cite{liu2019,lee2020,lee2021}, see section~\ref{sec:GPR}.
This feature selection also provides not only computation cost reduction but also a better physical understanding of the underlying PDEs.

\subsection{Machine Learning Regression Methods and Feature Selection}

\subsubsection{Gaussian Process Regression (GP) and Feature Selection via ARD}
\label{sec:GPR}
In Gaussian process regression, we describe probability distributions of target (hidden) functions under the premise that these target functions are sampled from a (hidden) Gaussian process, which is a collection of random variables such that the joint distribution of every finite subset of random variables is multivariate Guassian.
This assumption is reasonable when the target functions, 
$\frac{\partial b}{\partial t} = F\left(b,\frac{\partial b}{\partial x}, \frac{\partial^2 b}{\partial x^2}, \dots, s, \frac{ds}{dx}, \frac{d^2 s}{d x^2}, \dots   \right)$ 
are continuous and sufficiently smooth in a target domain~\cite{rasmussen2006, lee2019, lee2020}.
To identify the hidden Gaussian process from the data \{$\mathbf{x},y = F(\mathbf{x})=\frac{\partial b}{\partial t}$\} (where $\mathbf{x}$ represent the input vectors), we need to approximate a mean, $m(\mathbf{x})=\mathbb{E}[F(\mathbf{x})]$ and a covariance function, $K(\mathbf{x},\mathbf{x'})=\mathbb{E}[(F(\mathbf{x})-m(\mathbf{x}))(F(\mathbf{x'})-m(\mathbf{x'}))]$ from the given data:
\begin{equation}
    F \sim GP(m(\mathbf{x}), K(\mathbf{x},\mathbf{x'})).
\end{equation}
Generally, we assign a zero mean function for $m(\mathbf{x})$ and approximate the covariance function, $K(\mathbf{x},\mathbf{x'})$, based on the training data set.
%
% To approximate this covariance matrix, the Euclidean distance-based kernel function is used.
%
To approximate the covariance matrix, we employ a radial basis kernel function (RBF, denoted $\kappa(\cdot, \cdot)$), which is the 
%\emph{de facto} 
default kernel function in Gaussian process regression:
\begin{equation}
K_{ij}=\kappa(\mathbf{x_i},\mathbf{x_j};\theta) = \theta_0\exp \left( -\frac{1}{2} \frac{(\mathbf{x_{i}} - \mathbf{x_{j}})^2}{\theta_1} \right),
\label{eqn:kernel}
\end{equation} 
$\theta = [\theta_0, \theta_1]^T$ is a vector of hyperparameters to be optimized. 
The optimal hyperparameter set $\theta^*$ can be obtained by minimizing a negative log marginal likelihood over the training data set (of $n_{\mathrm{train}}$ data points). Using the kernel function with optimal hyperparameters, we represent a ($n_{\mathrm{train}}+n_{\mathrm{test}}$)-dimensional Gaussian distribution as 
\begin{equation}
\begin{bmatrix} 
\mathbf{y} \\
\mathbf{y^*} 
\end{bmatrix}
\sim
N \left( 
\mathbf{0},
\begin{bmatrix} 
\mathbf{K} + \sigma^2 \mathbf{I} &  \mathbf{K}_*\\
\mathbf{K}_*^T & \mathbf{K}_{**} 
\end{bmatrix}
\right),
\end{equation}
where $\mathbf{y^*}$ is our prediction of the distribution at the test conditions, $\mathbf{x^*}$, $\mathbf{K}_*$ represents the covariance matrix between training and test data ,while $\mathbf{K}_{**}$ represents the covariance matrix between test data. Also, $\sigma^2$ and $\mathbf{I}$ represent the variance of the (Gaussian) observation noise and the $n \times n$ identity matrix, respectively.

The predicted distribution (posterior) of $n_{\mathrm{test}}$ data, $\mathbf{y}^*$, is a Gaussian distribution conditioned on the training data:
\begin{equation}
\mathbf{{y}^*} \sim N\left(\mathbf{K}_*(\mathbf{K}+\sigma^2\mathbf{I})^{-1}\mathbf{y}, \mathbf{K}_{**} - \mathbf{K}_*^T(\mathbf{K}+\sigma^2\mathbf{I})^{-1}\mathbf{K}_* \right);
\end{equation}
we use the predicted mean ($\mathbf{\bar{y}^*}$) as an estimate of our Quantity of Interest, e.g. 
of the local time derivative of the bacterial density.

To identify the salient input features (feature selection), we modify the kernel function from the  default in Eq.(\ref{eqn:kernel}).
The ARD modification consists of assigning
an individual hyperparameter $\theta_l$ to each input feature (dimension) in the new covariance kernel, leading to a higher overall dimensional hyperparameter set as

\begin{equation}
K_{ij}=\kappa(\mathbf{x_i},\mathbf{x_j};\theta) = \theta_0\exp \left( -\frac{1}{2} \sum_{l=1}^k \frac{(x_{i,l} - x_{j,l})^2}{\theta_l
} \right),
\label{eqn:kernelard}
\end{equation} 
where $\theta = [\theta_0, \dots, \theta_k]^T$ is a $k+1$ dimensional vector of hyperparameters and $k$ is the number of dimensions of the input data domain.

After hyperparameter optimization, we check the magnitude of the optimal hyperparameters.
As shown in Eq.(\ref{eqn:kernelard}), a large magnitude of the hyperparameter $\theta_l$ nullifies the contribution along that direction to the distance metric (in the numerator), leading to
the relative insignificance (low sensitivity) of that input feature towards the predictions. 
Hence, we select only a few input features that have ``relatively" small magnitudes of the hyperparameters $\theta_i$.
After selecting a few salient features, we reconstruct the reduced GP model with only these selected input features, resulting in computational cost reduction.
In our case this will lead to learning an approximate equation using {\em less overall spatial derivatives
in the Right-Hand-Side} (RHS) than the ones retained in the physical model.

\subsubsection{Feedforward neural networks (FNN)}

Here, for learning the right-hand-side of the effective PDE, we have used an FNN 
with two hidden layers and a linear output layer. For each point in space and time $(x,t)$, the input to the FNN consists of the values of the cell density at the point, $b(x,t)$, its spatial derivatives, the chemoattractant profile $s(x)$ and its spatial derivatives, say, $\boldsymbol{u}=[b,\frac{\partial b}{\partial x}, \frac{\partial^2 b}{\partial x^2}, \dots, \frac{\partial^m b}{\partial x^m}, s, \frac{d s}{d x}, \frac{d^2s}{d x^2},\dots, \frac{d^k s}{d x^k}]^{T} \in\mathbb{R}^{k+m+2}$; while its output $F(\boldsymbol{u}): \mathbb{R}^{m+k+2}\rightarrow \mathbb{R}$ is the associated time derivative $\frac{\partial b(x,t)}{\partial t}$ {\em at the same $(x,t)$ point}. Thus, the FNN reads:
\begin{equation}
   F(\boldsymbol{u}) = \boldsymbol{W}^{o}\boldsymbol{\Phi}_2( \boldsymbol{W}_2\boldsymbol{\Phi}_1(\boldsymbol{W}_1\boldsymbol{u}+\boldsymbol{\beta}_1)+\boldsymbol{\beta}_2)+\beta_0.
\end{equation}
$\mathbf{W^{o}}$ is the matrix containing the weights connecting the second hidden layer to the linear output layer, $\boldsymbol{\Phi}_1, \boldsymbol{\Phi}_2$ denote the activation functions of the first and second hidden layers respectively, $\boldsymbol{W}_1$ is the matrix containing the weights from the input to the first hidden layer, $\boldsymbol{W}_2$ is the matrix with the weights connecting the first hidden to the second hidden layer, $\boldsymbol{\beta}_1$, $\boldsymbol{\beta}_2$ are the vectors containing the biases of the nodes in the first and second layers, respectively, and $\beta_0$ is the bias of the output node.
As has been proved by Chen and Chen \cite{chen1995universal}, such a structure (with sufficient neurons) can approximate, to any accuracy, non-linear laws of the time evolution of dynamical systems.
\subsection{Extraction of coarse-scale PDEs from the microscopic Monte Carlo simulations}
We learn three different types of data-driven PDE models, namely, (1) a black-box, (2) a gray-box, and, (3) PDE RHS's whose closures are learned (based on GPs and FNNs) {\em as corrections} of the analytically available Keller-Segel PDE closure; how many spatial derivatives are kept in the identified RHS relies on the ARD process above. 

In what follows, we describe the basic steps of our numerical scheme.
\begin{figure}[!htp]
  \centering
  \includegraphics[scale=0.5]{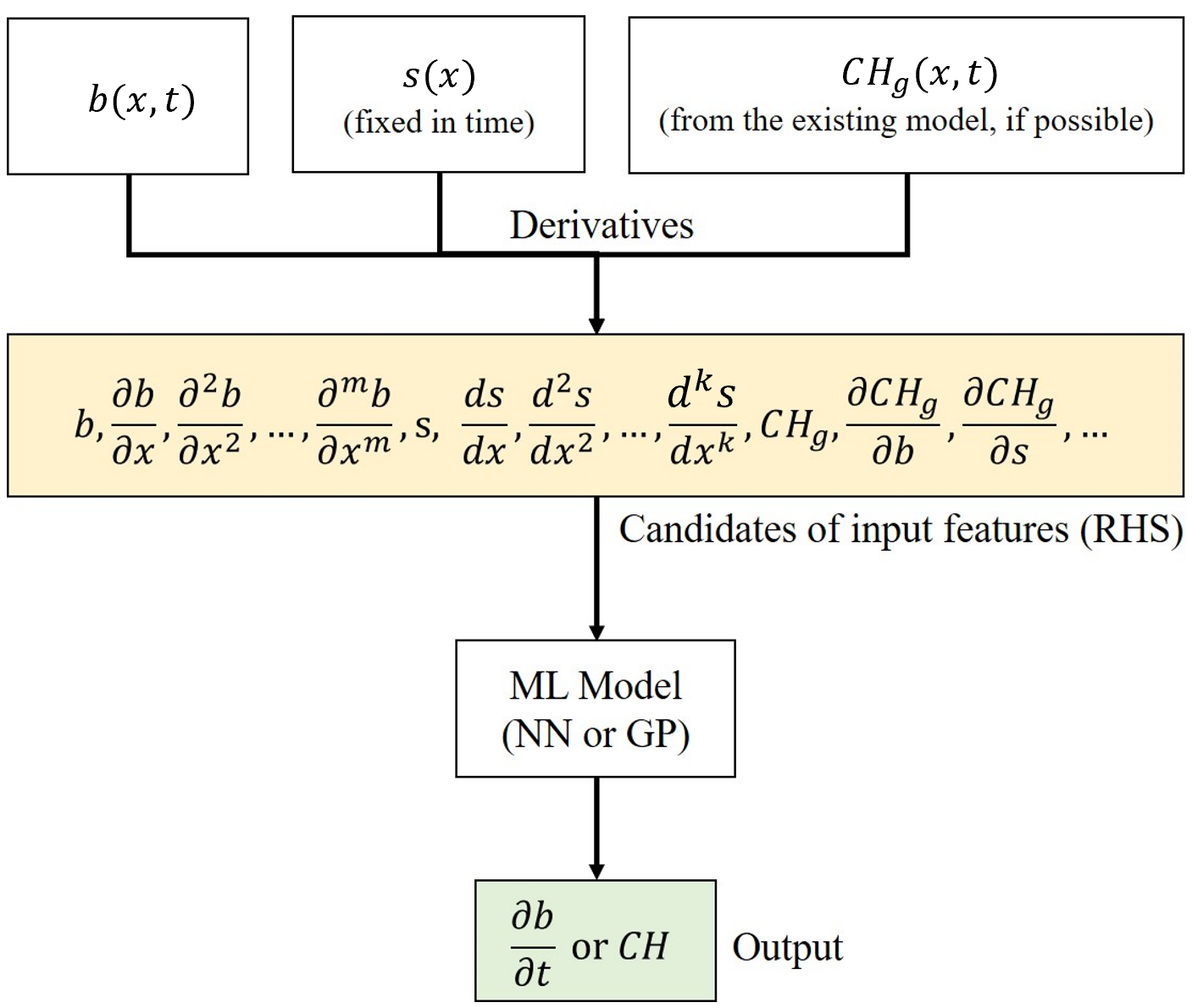}
  \caption{A schematic of the proposed data-driven numerical methodology: given the bacterial density profile in space and time, as well as the chemoattractant profile, (and, for the closure correction approach, the analytically available approximate chemotactic term $CH_g$) from Eq.(\ref{eqn:macroPDEOthmer})
 and their spatial derivatives, we extract possible reduced candidates as input features for learning
 the PDE right-hand-side (RHS) operator. Then, we train our machine learning models (NN or GP) to approximate $\frac{\partial b}{\partial t}$ or $CH$ (the ``ground truth" chemotactic term, see \ref{subsubsec:gray}).}
  \label{fig:schematic}
\end{figure}

%%%%%%%%%%%%%% BLACK BOX
\subsubsection{Black-box model}
We start by learning a black-box model for the local dependence of the time derivative of the bacterial density, $\frac{\partial b}{\partial t}$,
on the (local) density and its spatial derivatives as well as on the local chemoattractant concentration
and its derivatives, (i.e. the chemotactic PDE RHS operator) as:

\begin{equation}
\frac{\partial b}{\partial t} = B\left(b,\frac{\partial b}{\partial x}, \frac{\partial^2 b}{\partial x^2}, \dots, \frac{\partial^m b}{\partial x^m}, s, \frac{ds}{dx}, \frac{d^2s}{dx^2},\dots, \frac{d^ks}{dx^k}\right),
\end{equation}
The orders $m$ and $k$ are identified here by the ARD feature selection algorithm (see Table \ref{tab:features}).

\subsubsection{Gray-box model}
\label{subsubsec:gray}
In several cases, we may know some component of the macroscopic dynamic behavior that comes from intuition, previous studies and/or experiments. For example, one can compute through simulations and/or the aid of statistical mechanics the diffusion coefficient of the bacterial density.
One can then write a gray-box model which contains a Known Term --such as the diffusion term-- $KT$, and Unknown Terms (the chemotactic terms) $UT$:
\begin{equation}
\frac{\partial b}{\partial t} = UT\left(b,\frac{\partial b}{\partial x}, \frac{\partial^2 b}{\partial x^2}, \dots, \frac{\partial^m b}{\partial x^m}, s, \frac{ds}{dx}, \frac{d^2s}{dx^2},\dots, \frac{d^ks}{dx^k}\right) + KT. 
\end{equation}
Here, we assume that we know (or we can infer) the diffusion term, including the corresponding diffusion coefficient, $D$, i.e., $KT_\mathrm{Known} = D\frac{\partial^2 b}{\partial x^2}$.
In fact, here $D$ is computed from appropriately designed microscopic random-walk/Brownian motion Monte Carlo simulations {\em in the absence of a chemical gradient}, 
(see appendix~\ref{sec:diffusion} for more details). Therefore, in our case, the Unknown Terms include just the chemotactic term, i.e. $\frac{\partial b}{\partial t} =   D\frac{\partial^2 b}{\partial x^2} + CH$ or $CH = \frac{\partial b}{\partial t} -  D\frac{\partial^2 b}{\partial x^2}$.

\subsubsection{Learning closures and their corrections}
If some partial information is known (e.g. some of the terms in the RHS of the PDE), we can apply the gray-box approach discussed. 
However, sometimes, we may have a closed form, analytical, qualitatively good (but less accurate quantitatively) PDE model/closure, capable of describing {\em in a qualitative manner} similar macro-scale dynamics.
For example, we may have a closed-form PDE RHS that is
capable of predicting qualitative trends, but is not suitable for our particular experimental setup (e.g., one obtained for different types of chemoattractants, different types of bacteria, etc).
This can be thought of as a ``low-fidelity" model and
it can be used in the same spirit it would be used in a  {\em multifidelity} data fusion context (see~\cite{lee2019, Perdikaris_2017}).
Exploiting such existing approximate models, we propose a different type of gray-box machine learning scheme to calibrate the model to observation data, so 
as to match our specific experimental set-up/our detailed numerical simulations.

We propose four types of data-driven closure corrections to enhance the accuracy of an effective ``low-fidelity" PDE. In particular, we used the  ``generalized PDE" model for the chemotactic dynamics introduced in~\cite{erban2004} (please refer to section~\ref{sec:othmer} for more details) as the
low-fidelity reference model we want to correct. The four types of closure correction models are detailed in the schematic Figure~\ref{fig:schematic} and in Table~\ref{tab:features}.
First, we used a machine-learning scheme (based on GP or an FNN), to correct {\em the chemotactic term} $CH_g$ of the generalized PDE  so as to learn the {\em true} chemotatic term $CH$.
It may be that the quantitative closure is a simple, smooth function $F$ of the analytical approximate closure $CH_g$ in the general form of:
\begin{equation}
    CH(b,\frac{\partial b}{\partial x}, \dots, \frac{\partial^m b}{\partial x^m}, s, \frac{d s}{d x},\dots,\frac{d^k s}{d k^2}) = F(CH_g (b,\frac{\partial b}{\partial x},\dots, \frac{\partial^m b}{\partial x^m}, s, \frac{d s}{dx},\dots \frac{d^k s}{d k^2})).
\end{equation}
More often than not, this does not suffice, and more
information/more variables are necessary for quantitative prediction. 
Our first approach is to exploit data-driven embedding theories (in the spirit of Whitney/Takens embeddings ~\cite{Whitney36, Nash66, Takens81, lee2019})
to discover corrections from the {\em known} $CH_g$ to the {\em unknown} $CH$ using the first functional derivatives of $CH_g$ wrt. its variables~\cite{lee2019}:
\begin{equation} \label{eqn:closure_model}
    CH = H(CH_g, \frac{\partial CH_g}{\partial b}, \frac{\partial CH_g}{\partial b_x}, \frac{\partial CH_g}{\partial s}, \frac{\partial CH_g}{\partial s_x}, \frac{\partial CH_g}{\partial s_{xx}}). 
\end{equation}
Within the GP framework, we also identified a second ``version" of this
closure correction, using fewer, dominant such derivatives via ARD analysis as:
\begin{equation} \label{eqn:closure_model_red}
     CH = H(\frac{\partial CH_g}{\partial b}, \frac{\partial CH_g}{\partial b_x}, \frac{\partial CH_g}{\partial s},\frac{\partial CH_g}{\partial s_x}, \frac{\partial CH_g}{\partial s_{xx}}). 
\end{equation}

As a second idea, also conceptually based on embedding theories, we also considered another closure correction approach, in which the equation RHS was not just a function of the approximate $CH_g$, but also included additional local inputs (in our
first attempt we included the local bacterial density $b$ and the local
chemoattractant $s$) as additional information, so that the corrected
closure is a learned function of the form 
\begin{equation}
     CH = h({CH_g}, b, s). 
\end{equation}

Finally, we also tried a simple ``additive" correction; we learned the bias term between the observed chemotactic term and the approximated chemotactic term of the generalized PDE in the form:
\begin{equation}
     CH - CH_g = H(b,\frac{\partial b}{\partial x},\frac{\partial^2 b}{\partial x^2}, s, \frac{ds}{dx}, \frac{d^2s}{dx^2}). 
\end{equation}

\begin{table}[!htp]
\label{tab:features}
\centering
\begin{tabular}{llc}
    \hline
    Data-driven Model & Input Features & Output \\ \hline
    Black-box (Full) & $b,\frac{\partial b}{\partial x}, \frac{\partial^2 b}{\partial x^2}, s, \frac{ds}{dx},\frac{d^2s}{dx^2}$  &  $\frac{\partial b}{\partial t}$ \\
    Black-box (Reduced)  & $b,\frac{\partial b}{\partial x}, \frac{\partial^2 b}{\partial x^2}, s, \frac{ds}{dx}$  &  $\frac{\partial b}{\partial t}$ \\
    Gray-box (Full)  & $b,\frac{\partial b}{\partial x}, \frac{\partial^2 b}{\partial x^2}, s, \frac{ds}{dx},\frac{d^2s}{dx^2}$  &  $CH$ \\
    Gray-box (Reduced)  & $b,\frac{\partial b}{\partial x}, \frac{\partial^2 b}{\partial x^2}, s, \frac{ds}{dx}$ &  $CH$ \\
    Functional Correction (Full) & $CH_g, \frac{\partial CH_g}{\partial b}, \frac{\partial CH_g}{\partial b_x}, \frac{\partial CH_g}{\partial s}, \frac{\partial CH_g}{\partial s_x}, \frac{\partial CH_g}{\partial s_{xx}}$ & $CH$ \\
    Functional Correction (Reduced) & $\frac{\partial CH_g}{\partial b}, \frac{\partial CH_g}{\partial b_x}, \frac{\partial CH_g}{\partial s}, \frac{\partial CH_g}{\partial s_x}, \frac{\partial CH_g}{\partial s_{xx}}$ & $CH$ \\
    Correction (No derivatives) & $CH_g, b, s$ & $CH$ \\
    Additive Correction & $b,\frac{\partial b}{\partial x}, \frac{\partial^2 b}{\partial x^2}, s, \frac{ds}{dx},\frac{d^2s}{dx^2}$ & $CH-CH_g$\\ \hline
\end{tabular}
\caption{Selected groups of input features and the corresponding predicted quantity (output) for different data-driven PDE law correction approaches. ``Reduced" represents models constructed using the (fewer) input features selected via ARD within the Gaussian process framework. We only reduce GP models through ARD; the corresponding NN reduction via autoencoders was not attempted. 
For the definitions of $CH, CH_g$ see Eq. (\ref{eqn:macroPDEOthmer}) and section \ref{subsubsec:gray} respectively}.
\end{table}

\section{Results}
\label{sec:models}

For our computations, we run a Monte Carlo simulation of $n=5000$ bacteria initially located at $x=5.5$, from $t=0$ to $t=5000s$ with $dt=2s$ as reporting horizon, and collect the training data.
For training, we collected data from four (fixed in time) different chemo-nutrient concentration profiles, all with a Gaussian distribution $s(x) = \frac{1}{\sqrt{2\pi\sigma^2}}\exp(-\frac{1}{2}\frac{(x-\mu)^2}{\sigma})$: (1) $\mu=6, \sigma=1$; (2) $\mu=6, \sigma=1.5$; (3) $\mu=7, \sigma=1.5$; (4) $\mu=7, \sigma=1.25$.
Specifically, from 5000 individual trajectories of the bacterial motion, we estimated the normalized bacterial density $b(x,t)$ on a uniform grid from $x=3$ to $x=9$ with $dx=0.05$ at every time step using kernel smoothing~\cite{Bowman1997} as:
\begin{equation}
    b(x,t) = \frac{1}{nh}\sum^{n}_{i=1}K\left(\frac{x-x_i(t)}{h}\right),
\end{equation}
where $K(\cdot)$ represents the kernel smoothing function (here, a Gaussian function), and $h$ is the bandwidth (here, set to $h=0.3$).
For the approximation of the first $\partial b/\partial x$ and $\partial b/ \partial t$ and the second $\partial^2b/\partial x^2$ partial derivatives of the density profile, we used %second-order
central finite differences.
Thus, our data-driven models are constructed based on six input features ($b$, $\partial b/ \partial x$, $\partial^2b/\partial x^2$, $s$, $ds/dx$ and $d^2s/dx^2$)  that are used for learning  the corresponding time derivative $\partial b/ \partial t$. 

Gaussian Process learning was performed in Matlab using a RBF kernel with ARD. For feature selection, the cut-off is $10^5$. That is, if the optimal hyperparameter value is higher than $10^5$, we eliminate the corresponding input features.
Neural Network learning was performed in Python using Tensorflow \cite{tensorflow2015-whitepaper} with two hidden layers with $[9,8,8]$ neurons each (Black box, Gray box, Correction model respectively), equipped with a hyperbolic tangent activation function. The Neural Network was trained using an Adam optimizer \cite{adam_optimizer} and the training hyperparameters were tuned empirically (Epochs [$2,560$, $10,240$, $2,560$],  Batches [$800,000$, $750,000$, $300,000$], Initial Learning Rate $0.02$ with a plateau learning rate scheduler with patience $1,200$ epochs and factor $0.5$).

\begin{figure}
  \makebox[\textwidth][c]{\includegraphics[width=2\textwidth]{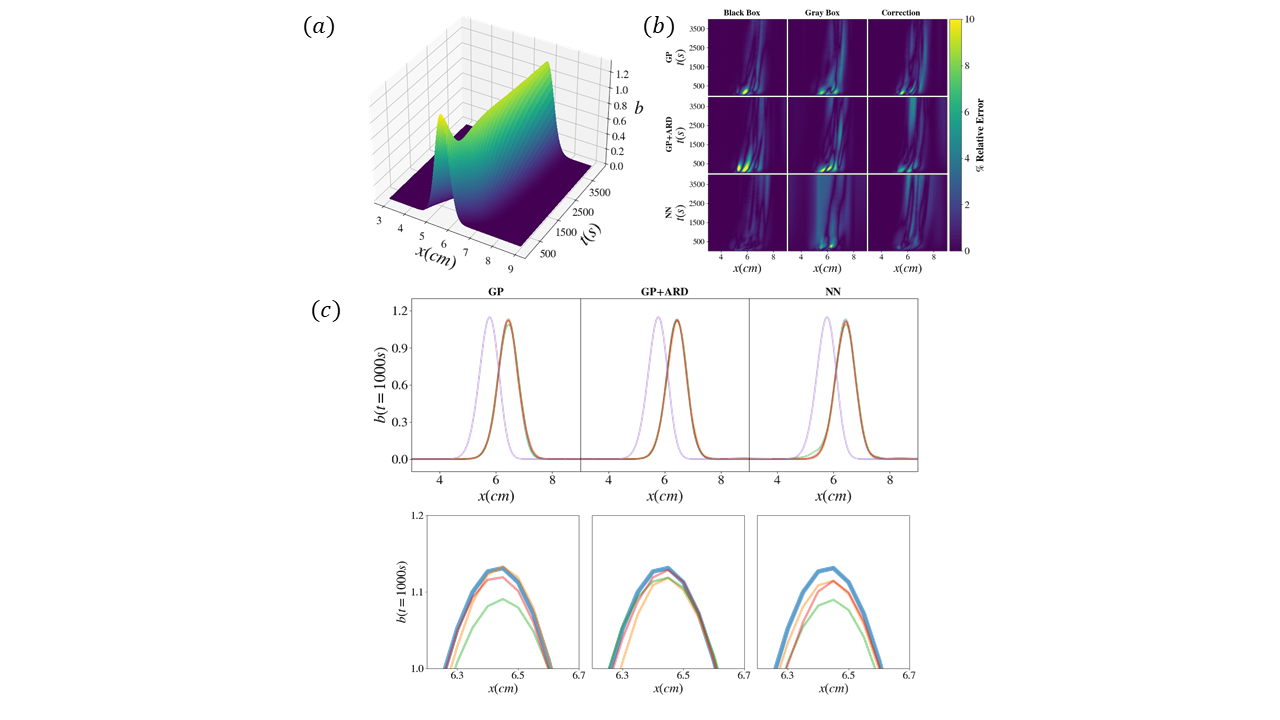}}%
  \caption{(a) \textbf{Training/reconstruction} case. Ground Truth (GT) evolution for $\mu = 7,  \sigma = 1.25$: smoothened profiles of bacterial density derived from post-processing agent-based simulations. (b) Quantitative performance of representative data-driven models (DD) trained by Gaussian Process Regression (GP), reduced Gaussian Process Regression with ARD (GP+ARD) or a Neural Network (NN): relative error (\%) based on maximum density for black-box model, gray-box model, and correction model (first, second and third rows respectively). (c) Qualitative comparison of profiles of bacterial density at $t=1000\mathrm{s}$: PDEs learned through (left) Gaussian process; (middle) Reduced Gaussian processes (with ARD); and (right) Neural Networks.Profiles are colored as follows: blue -- ground truth, orange -- black box, green -- gray box, purple -- PDE Eq.[3]. The bottom row is a blowup of the profile's peak. Note that for the Monte-Carlo simulations the initial state ($t=0\mathrm{s}$) is at $x=5.5\mathrm{cm}$ for all agents, while all PDE simulations (DD models, PDE in \ref{eqn:macroPDEOthmer}) begin at $t=20\mathrm{s}$. No flux boundary conditions were used, consistent with \cite{erban2004}.}  \label{fig:result_reconstruction}
\end{figure}

After learning the time evolution operator, we first tested whether the laws identified could reproduce the trajectories from which the training data were collected. For illustration, we performed numerical integration with the data-driven learned PDEs (using both GP and FNN) from $t=20\mathrm{s}$ to $t=4020\mathrm{s}$ with $dt=2\mathrm{s}$ using a 4-th order Runge-Kutta scheme, as well as 
a commercial package integrator (explicit Runge-Kutta method of order 5(4) \cite{DORMAND198019} as implemented by $solve\_ivp$ in Python, 
resulting in a maximum absolute error of $4\cdot 10^{-6}$ and the corresponding integrator (and tolerances) in Matlab with  maximum absolute error of $9\cdot 10^{-4}$.
High spatial frequency Fourier modes of the bacterial density profile 
were consistently filtered (a procedure analogous to adding hyperviscosity
in hydrodynamic models, \cite{Thiem_2021})
The ground truth spatiotemporal evolution of the bacterial density is shown in Figure~\ref{fig:result_reconstruction}(a) and the corresponding relative errors are shown in Figure~\ref{fig:result_reconstruction}(b) while in Figure~\ref{fig:result_reconstruction}(c), we show the reconstructed profile at $t=1000\mathrm{s}$ for one of our training chemo-nutrient profiles $s(x) = \frac{1}{\sqrt{2\pi 1.25^2}}\exp(-\frac{1}{2}\frac{(x-7)^2}{1.25^2})$)
The performance of our several different data-driven PDE closure corrections was assessed in terms of the 
relative approximation error between the ground truth density profiles ($b^{GT}(x,t)$)
and the profiles ($b^{DD}(x,t)$) resulting from numerical integration of the learned approximate PDE right-hand-sides.
This error was defined as
$E_r = 100\frac{|b^{GT}(x,t)-b^{DD}(x,t)|}{\max b^{GT}(x,t)}$;
%$E_r = 100\times\frac{b^{GT}(x,t)-b^{DD}(x,t)}{\max b^{GT}(x,t)}$;
a comparison to the density profile of the full Monte Carlo simulations at $t=1000\mathrm{s}$ is also provided.

\begin{figure}
  \makebox[\textwidth][c]{\includegraphics[width=2\textwidth]{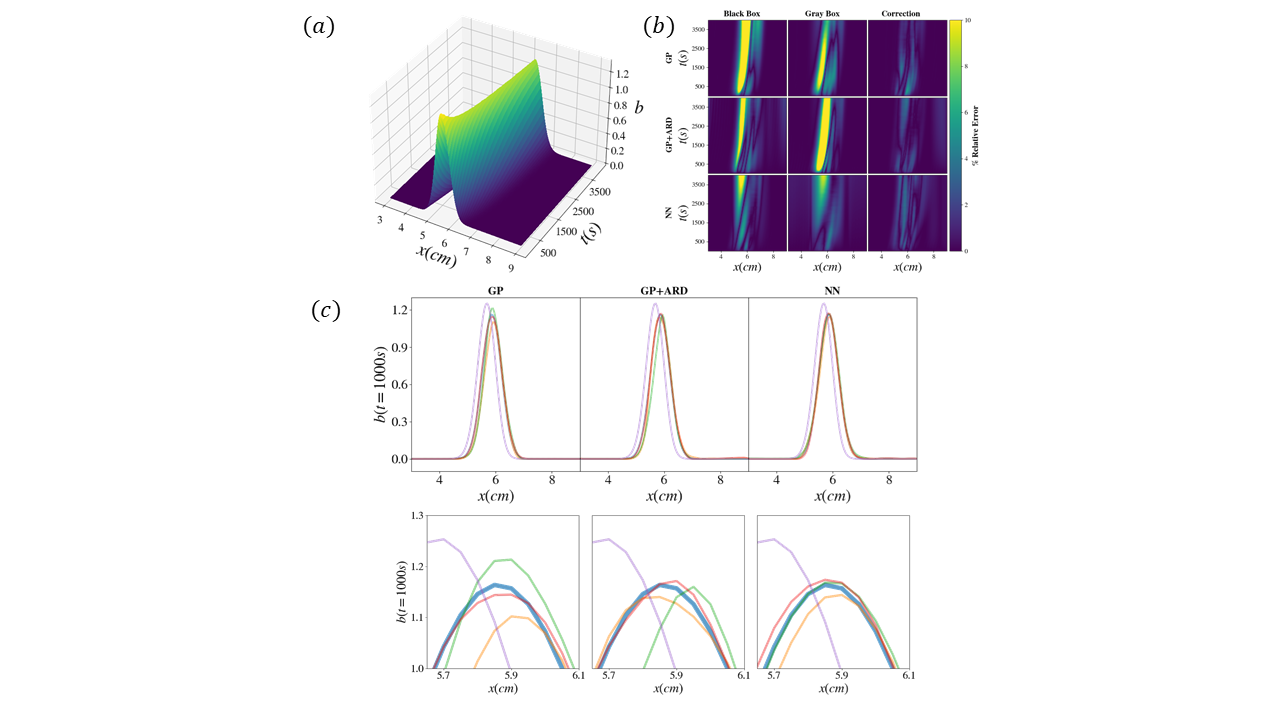}}%
  \caption{ (a) \textbf{Testing} case. Ground Truth (GT) evolution for $\mu = 6.5,  \sigma = 1.35$: smoothened profiles of bacterial density derived from post-processing agent-based simulations. (b) Quantitative performance of representative data-driven models (DD) trained by Gaussian Process Regression (GP), reduced Gaussian Process Regression with ARD (GP+ARD) or a Neural Network (NN): relative error (\%) based on maximum density for black-box model, gray-box model, and correction model (first, second and third rows respectively). (c) Qualitative comparison of profiles of bacterial density at $t=1000\mathrm{s}$: PDEs learned through (left) Gaussian process; (middle) Reduced Gaussian processes (with ARD); and (right) Neural Networks.Profiles are colored as follows: blue -- ground truth, orange -- black box, green -- gray box, purple -- PDE Eq.[3]. The bottom row is a blowup of the profile's peak. Note that for the Monte-Carlo simulations the initial state ($t=0\mathrm{s}$) is at $x=5.5\mathrm{cm}$ for all agents, while all PDE simulations (DD models, PDE in \ref{eqn:macroPDEOthmer}) begin at $t=20\mathrm{s}$. No flux boundary conditions were used, consistent with \cite{erban2004}.}
  \label{fig:result_test}
\end{figure}

After that, we tested the performance of the data-driven PDE closure corrections with the test data from a new chemoattractant concentration profile (not included in the training data set): $s(x) = \frac{1}{\sqrt{2\pi 1.35^2}}\exp(-\frac{1}{2}\frac{(x-6.5)^2}{1.35^2})$. The ground truth of the testing case is plotted in Figure~\ref{fig:result_test}(a).
The predicted profile at $t=1000\mathrm{s}$ and the corresponding relative errors are shown in Figures~\ref{fig:result_test}(b) and (c), respectively.
Table~\ref{tab:features}, summarizes the different data-driven models with respect to (1) machine learning techniques, (2) selected features, and (3) the corresponding predicted quantity.

A benefit of reduced feature selection is that the computational cost in the training phase is reduced, while (as shown in Figures~\ref{fig:result_reconstruction} and ~\ref{fig:result_test}) the predictive accuracy is retained. 
Regarding the black-box trajectory reconstruction relative error for the training data set, the GP data-driven models never exceed $14\%$ relative error when integrated for a long time ($15\%$ for the ARD-reduced GP), while the FNN data-driven models never exceed $4\%$ relative error.
For the test data-set, the respective maximum relative errors are $20\%$ and $12\%$ for the GP (full or reduced) and FNN, respectively. 

Interestingly, the largest errors observed for the GP are concentrated during the fast, initial transient, while the trajectories become more accurate at later times as they approach steady state. FNNs seem to perform better in capturing this initial transient.

As shown in Figures~\ref{fig:result_reconstruction} and ~\ref{fig:result_test}, the gray-box models provide comparable, yet slightly improved accuracy, compared to the black-box ones. 
Specifically, the maximum reconstruction relative errors for the training data set are $9\%$ for the GP ($12\%$ for reduced GP) and $10\%$ for the FNN models, while for the testing case these are $12\%$ for the GP ($19\%$ for reduced GP) and $10\%$ for the FNN models.
These results confirm the capabilities of gray-box models, which combine partial physical knowledge (exact, or even approximate) with data-driven information towards
accurate and efficient data-assisted modelling of complex systems. 

There are, of course no guarantees here for the accurate generalization of the predictions beyond the training data; yet the performance of our models over the test set, and also for chemoattractant profiles not included in the training,
appears promising.
Our expectation is that the closure correction models, by always making use of the ``low-fidelity" (qualitative) information at every time step, may generalize better.

Here, results are presented for two of the four closure correction approaches , i.e. the ones described in Eq.(\ref{eqn:closure_model}, \ref{eqn:closure_model_red}). In particular, the maximum  relative errors for the training case were $10\%$ for the GP ($8\%$ for reduced GP) and $6\%$ for the FNN models, while for the testing case they were $5\%$ for the GP ($4\%$ for reduced GP) and $4\%$ for the FNN models. 
Thus, all different closure correction approaches (even though the figures for the
``additive" and the ``no derivatives" corrections are not included for economy of space), provide reasonable accuracy for validation as well as test profiles qualitatively and even quantitatively.

\section{Conclusions}
\label{sec:conclusion}
Machine learning has long been used to solve the inverse problem, i.e. the identification of nonlinear dynamical systems models from data \cite{Rico1992,Krischer93,Masri,Rico94,chen1995universal,Gonzalez98,siettos2002truncated,siettos2002semiglobal,alexandridis2002modelling}. Recently, due to technological and theoretical advances, there has been a renewed interest, especially for the case of multiscale/complex systems \cite{raissi2019physics,kemeth2020learning,lee2020,vlachas2020backpropagation,chen2021solving,karniadakis2021physics}.
We presented a machine-learning framework for the numerical solution of the inverse problem in chemotaxis. In particular, we showed how one can learn black-box and gray-box parabolic PDEs for the emergent dynamics of bacterial density evolution (and, importantly, unknown closures and their corrections) directly from high-fidelity microscopic/stochastic simulations.
Specifically, we introduced a computational data-driven framework for nonlinear PDE/closure identification and correction; the framework 
consisted of three progressively more
physics-informed processes:
(a) learning a black-box PDE (learning the right-hand-side of an coarse-scale PDE including the diffusion term), (b) learning a gray box PDE (an entire unknown closure, 
with a known Diffusion term), and (c) obtaining closure corrections (providing a correction
of an analytically available closure in a low-fidelity, approximate PDE model).
Within this framework, we exploited the Automatic Relevance Determination (ARD) algorithm for feature selection, in order to
reduce the number of variables on which
the closure depends. 
We note that a discussion of the possibility (in the spirit of 
Takens embeddings) of using short term measurement histories to mitigate partial observations is the subject of a forthcoming manuscript. \cite{Psarellis_upcoming}. 
Our overall approach forms a bridge between analytical/mechanistic/physical understanding, and data-driven ``black-box” or ``gray-box" learning of physical process dynamics, allowing for a synergy between varying types of physical terms/models and data-driven terms/models.

{\bf Acknowledgements}  This work was partially supported by the US Department of Energy, by the US Air Force Office of Scientific Research and by DARPA. C. S.  was partially supported by INdAM, through GNCS and the Italian research fund FISR2020IP - 02893.

\appendix

\section{Details on the Monte Carlo chemotaxis model}

Our microscopic, agent-based model is based on the work of \cite{othmer1998oscillatory, othmer2013excitation}.
Each bacterium is modelled as having six flagellae; 
special care has been taken in modelling the direction of the rotation of the flagellar filaments, as this constitutes the basis of chemotaxis \cite{larsen1974change,spiro1997}.
Following \cite{scharf1998control}, the motor dynamics are described by a two-state system modelling the transition rates (transition probabilities per unit time) between CCW and CW (counter-clockwise and clockwise, respectively) rotation for each flagellum. These are characterized by an exponential distribution of time intervals in each state \cite{turner1996temperature}. Let us denote by $k^+$ ($k^-$) the transition rate from CCW to CW (CW to CCW). Then, the bias of CW, i.e. the fraction of time that a flagellum rotates CW is $p_{CW}=\frac{k^+}{k^{+}+k^-}$, $p_{CCW}=1-p_{CW}$. The reversal frequency in the direction of rotation of the flagellar motors is  \cite{turner1996temperature}:
\begin{equation}
    \rho=p_{CCW} k_{+} + p_{CW} k_{-} = \frac{2k^+k^-}{k^++k^-}.
\end{equation}
and the rate constants are given by:
\begin{equation}
    k_{+}=\rho/(2 p_{CCW}),  k_{-}=\rho/(2 p_{CW})
\end{equation}
For a cell with $N$ flagellae, the total CCW bias of the cell is given by \cite{spiro1997}:
\begin{equation}
    P_{CCW} = \sum_{j=\theta}^{N} \binom{N}{j} p_{CCW}^{j}(1-p_{CCW})^{N-j}. 
    \label{eqn:voting}
\end{equation}
For $N=6$, $p_{CCW}=0.64$ and $\theta=N/2$, we get $P_{CCW}\sim 0.87$ suggesting that the cell spends around 90\% of the time running \cite{spiro1997}. 
This result is in line with experimental observations for the motility of wild-type E.coli in the absence of changes of the substrate, where the mean run (swimming) periods are $\sim$ 1s and the tumble periods $\sim$ 0.1 s (for the strain AW405 in dilute phosphate
buffer at $32^o C$) \cite{ishihara1983coordination}.

Experimental studies have shown that these rates depend on the CheY-P concentration, say $C$. In particular, Cluzel et al. \cite{cluzel2000ultrasensitive} have shown that the dependence of CW bias (between the values 0.1 and 0.9) to $C$ can be approximated by a Hill function with a coefficient  $H$ $\sim$ 10.3 $\pm$  1.1, with
a dissociation constant $K_d$= 3.1 mM/s. Thus, the CW bias reads:  
\begin{equation}
    p_{CW} = \frac{C^H}{K_{d}^H +C^H}.
    \label{eqn:CWbias}
\end{equation}
Based on the above findings, the transition rates $k^+$, $k^-$ are given by \cite{setayeshgar2005application}:
\begin{equation}
    k^+ = \frac{H C^{H-1}}{K_{d}^H + C^H},
\end{equation}
\begin{equation}
    k^- = \frac{1}{C} \frac{H K_{d}^{H}}{K_{d}^H + C^H}.
\end{equation}
Thus, based on the model formulation and nominal values of the parameters, the expected fraction of time spent in the CCW state in the absence of stimulus for each cell from kMC simulations is $\sim$ 0.855, close enough to the one observed experimentally. 
In the absence of spatial variations in the chemoattractant (or repellent) profile, the rotation of the flagellar filament is biased  towards the CCW direction (that is, the probability of CCW rotation of a flagellum is higher than that of CW rotation),
when viewed along the helix axis towards the point of insertion in the cell \cite{larsen1974change}. This bias depends on the type of bacterial strain and the temperature; for the wild-type strain AW405, it has been found that the average value of the CCW bias is 0.64 at $32^o C$ \cite{larsen1974change,BLOCK1982215}. When the majority of the flagellar filaments rotate CCW (CW) the cell swims (tumbles).

\begin{algorithm}%[H]
\label{alg:agent}
\SetAlgoLined
\textbf{{Initialize}} {Set positions $x(t=0)$, velocities $v(t=0)$ for all bacteria, the rotation direction of all flagellae, the states of the cartoon model $u_1(t=0), u_2(t=0)$ and the base sampling time $dt$.\\
\textbf{{Update}} positions $x(t+dt)$ of the bacteria at the next time step $(t+dt)$ as follows.}\\
Draw a random number $r$ from a uniform distribution in [0,1]. \\
For each one of the 6 flagellae, compute the probabilities of switching the direction of the rotation, i.e. $k_{+} \cdot dt$ if the flagellum rotates CCW and  $k_{-} \cdot dt$ if the flagellum rotates CW. If $k_{+} \cdot dt>1$ or $k_{+} \cdot dt>1$ halve the base sampling time.\\
\textbf{For} each one of the flagellae of each bacterium compare $r$ with $k_{\pm} \cdot dt$. \\
 \eIf{$r$ > $k_{\pm} \cdot dt$}{keep the same rotating direction}{change rotating direction}
 \eIf{the number of flagella that rotate CW > 3}{tumble}{\eIf{flagella previously was running}{continue to run to the same direction}{start running to the left ($dir=-1$) or to the right ($dir=+1$) with equal probability} }
 $x(t+dt) = x(t) + dir \cdot v \cdot dt$\\
 $u_1(t+dt) = u_1(t) + dt \cdot \frac{f(s)-u_1(t)-u_2(t)}{t_e}$\\
 $u_2(t+dt) = u_2(t) + dt \cdot \frac{f(s)-u_2(t)}{t_a}$\\
 \caption{Monte Carlo Model}
\end{algorithm}

\section{Determination of the parameters of the macroscopic PDE for bacterial density evolution.}

\subsection{Determination of the Diffusion Coefficient}
\label{sec:diffusion}
\begin{figure}[!htp]
  \centering
  \includegraphics[scale=0.5]{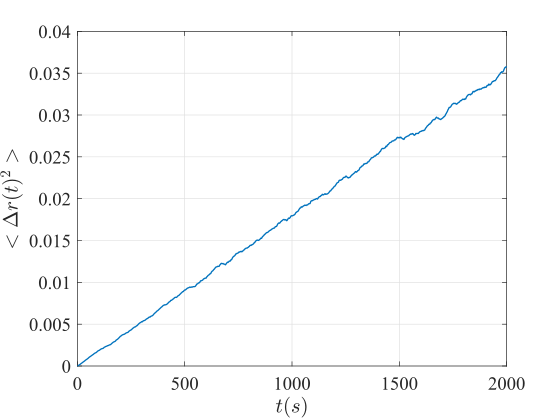}
  \caption{The average square displacement (over all 1000 cells) as a function of time in the absence of stimulus (gradient of chemoattractant), when all cells are initialized in the tumbling phase with $u_1(0)=0$, i.e. without excitation and fully adapted, i.e. $u_2(0)=f(s)$, and $v=0.003 cm/s$.}
  \label{fig:diffusion}
\end{figure}

An estimation of the diffusion coefficient for cell motility in the absence of stimulus can be attempted following two paths.
From a microscopic point of view, considering a random walk simulation, the mean free path i.e., the swimming distance without any change in the direction is given by $\delta r = \tau\cdot v$, where $\tau$ is the mean time of swimming in one direction. Considering $n$ such time steps in time $t$ (i.e. $n=t/\tau$), the total mean-squared displacement $\Delta r (t)^2$ at a certain time ($t$) is given by the Einstein relation \cite{berg1990chemotaxis}: 
\begin{equation}\label{Einstein}
   <\Delta r^2(t)>=2 D_m t \approx 2 n \delta r^2  = 2 \tau v^2 t,
\end{equation}
which is valid for $t>> \tau$, where, $\tau$ is the characteristic time scale.
%Thus, the diffusion coefficient for  $t>> \tau$ is given by $ D_m = \tau v^2$ Thus, for $v=0.003 cm/s$ and $\tau \sim 1 s$ \cite{berg1972chemotaxis,spiro1997,cluzel2000ultrasensitive}, we get $D_m \sim 9E-06 cm^2/s$. 
Here,  the value of $D_m$ is estimated from our Monte Carlo simulations in the absence of stimulus (we have set $s(x)=1$, $\forall x$), by tracking the trajectories of 1000 cells for a time period of $2000\mathrm{s}$. The cells are initially positioned at the middle of the domain, all initialized at the tumbling phase, with $u_1(0)=0$ (no excitation), and adapted

with $u_2(0)=f(s(x))$, with a constant velocity of $v=0.003 cm/s$ (as in \cite{berg1990chemotaxis}). Figure (\ref{fig:diffusion}) depicts the average of the square distance as a function of time. By least-squares, we get $\hat{D}_m \approx 9 \cdot 10^{-6} cm^2/s$. This is in good agreement with experimental observations for the E. Coli motility (see \cite{berg1972chemotaxis,berg1990chemotaxis,spiro1997,cluzel2000ultrasensitive}) 

From a macroscopic point of view, one can estimate the diffusion coefficient $D_M$ from a linear curve fitting between $\frac{\partial b}{\partial t}$ and $\frac{\partial^2 b}{\partial x^2}$ with finite difference approximations of temporal and spatial derivatives at the coarse-scale.
Thus, by fixing a spatial gradient of chemo-nutrient profile to zero ($\nabla c = 0$), we can consider a simple diffusion equation with a constant diffusion coefficient, $D$:
\begin{equation}
    \frac{\partial b}{\partial t} = D \nabla^2 b.
\end{equation}

Finally, we note that the Einstein relation for the diffusion coefficient given by Eq.(\ref{Einstein}) can be approximated on average over a run and tumble period as:
\begin{equation}\label{Einsteinapprox}
   <\Delta r^2>\approx\bar{v}^2 \bar{T}_{run}^2 =2 \bar{D}_m (\bar{T}_{run}+\bar{T}_{tumb}),
\end{equation}
where ${\bar{T}_{run}, \bar{T}_{tumb}}$ denote the average duration of swimming and tumbling periods, respectively, and $\bar{v}$ is the average swimming speed. Thus, based on Eq.(\ref{Einsteinapprox}) and assuming that the tumbling duration is negligible compared to the swimming duration (as assumed for the derivation of the generalized Keller-Segel theory embodied in Eq.(\ref{eqn:macroPDEOthmer})), an approximation of the diffusion coefficient is given by:
\begin{equation}\label{Difapprox}
   \bar{D}_m=\frac{\bar{v}^2}{2\lambda_0}, \qquad \lambda_0=\bar{T}_{run}^{-1}.
\end{equation}
Hence, setting $\bar{D}_m=\hat{D}_m\approx 9 \cdot 10^{-6}  cm^2/s$, $\lambda_0=1 s^{-1}$ (in agreement with experimental findings), we get $\bar{v}=\sqrt{2}v=3 \sqrt{2} \cdot 9 \cdot 10^{-3} cm/s$ as the average
velocity appearing in Eq.(\ref{eqn:macroPDEOthmer}).
\subsection{Determination of the parameter $c$ of the macroscopic PDE}
As stated in section 2, one of the assumptions for the derivation of the closed-form Keller-Segel equation (\ref{eqn:macroPDEOthmer}) is the linear relation between the turning frequency $\lambda$, and the basal frequency $\lambda_0 \sim 1 s^{-1}$, i.e. for each cell at position $x$ at time $t$, we have (see Eq.(\ref{eqn:lambda})):
\begin{equation}
   \lambda(x,t)=\lambda_0-c u_1(x,t).
\end{equation}
For initial values $u_1(x,0)$, $u_2(x,0)$ for all cells (i.e. $\forall x \in \rm I\!R$), the analytical solution of the cartoon model (Eq.\ref{eqn:cartoon1}) is given by:
\begin{multline}
   u_1(x,t)= \frac{{\mathrm{e}}^{-\frac{t}{t_{e}}}\,\left({K_{s}}^2\,t_{a}-{K_{s}}^2\,t_{e}+s^2\,t_{a}-s^2\,t_{e}+2\,K_{s}\,s\,t_{a}-2\,K_{s}\,s\,t_{e}\right)}{{\left(K_{s}+s\right)}^2\,\left(t_{a}-t_{e}\right)}\,u_{1}(0,x)+ \\
   \frac{{\mathrm{e}}^{-\frac{t}{t_{e}}}\,(s^2\,t_{a}\,u_{2}-k\,s\,t_{a}+{K_{s}}^2\,t_{a}\,u_{2}(0,x)+2\,K_{s}\,s\,t_{a}\,u_{2}(0,x))+k\,s\,t_{a}\,}{{\left(K_{s}+s\right)}^2\,\left(t_{a}-t_{e}\right)}- \\
   \frac{t_{a}\,{\mathrm{e}}^{-\frac{t}{t_{a}}}\,({K_{s}}^2\,u_{2}(0,x)-k\,s+s^2\,u_{2}(0,x)+2\,K_{s}\,s\,u_{2})+ k\,s\,t_a}{{\left(K_{s}+s\right)}^2\,\left(t_{a}-t_{e}\right)},
\end{multline}
\begin{equation}
   u_2(x,t)= \frac{{\mathrm{e}}^{-\frac{t}{t_{a}}}\,}{{\left(K_{s}+s\right)}}\,u_{2}(0,x)-\frac{k\,s({\mathrm{e}}^{-\frac{t}{t_{a}}}-1)}{{\left(K_{s}+s\right)}^2}.
   \label{eqn:analu2}
\end{equation}
Note that, if one sets as initial value $u_2(x,0)=f(s)$, then the second equation of the cartoon model (see Eq.(\ref{eqn:cartoon1})) gives $u_2(x,t)=f(s)$, $\forall t$ and the analytical solution for $u_1(x,t)$ is reduced to:
\begin{equation}
    u_1(x,t)=u_{1}(x,0){\mathrm{e}}^{-t/t_a}.
\end{equation}
To this end, the parameter $c$ in Eq.(\ref{eqn:lambda}) appearing in the Keller-Segel-class PDE given by Eq.(\ref{eqn:macroPDEOthmer}) can be found with the aid of Monte Carlo simulations, by fixing $u_{1}(x,t)$, $\forall t$ to different relatively small values, say $u_1$  $\forall x$, and  measuring,  the number of turning events $\lambda(u_1)$; then the value of the parameter $c$ can be estimated by least-squares. A different way would be to set an initial value for $u_1(x,0)$ (setting also as initial value $u_2(x,0)=f(s)$), run the Monte Carlo simulator, measure the turning frequencies $\lambda(u_1(t))$ and based on the above, the value of $c$ can be again estimated with least-squares.\par
Here, to estimate $c$, we have fixed $u_1$ to the following values: -0.02, -0.015, -0.01, -0.005, 0, 0.005, 0.01, 0.015, 0.02, where the linear relation between $\lambda$ and $\lambda_0$ is valid, and we computed $\lambda(u_1)$ based on Monte Carlo simulations with 1000 cells for a time period of 2000s. For these values, $\hat{\lambda_0}\sim 1 s^{-1}$ (in a good agreement with the experimental findings) and $\hat{c}\sim 19.5$ (99\% CI: 18-21). We note that this value is consistent with what has been reported in other studies \cite{xue2015macroscopic}. For our simulations with the macroscopic PDE, we have set $\hat{c}=20$.

\bibliography{Chemotaxis}

\end{document}